# Replicability and the Evolution of Scientific Norms

Hope Bretscher[1] and Núria Muñoz Garganté[2]

[1]Max Planck Institute for the Structure and Dynamics of Matter, Hamburg Germany
[2]Max Planck Institute for the History of Science, Berlin, Germany

*To avoid a replication crisis in physics, physicists need to understand how ever-changing social forces shape scientific practice — and even the underlying notions of replicability and objectivity.*

Recent years have seen high-profile retractions of controversial claims in physics — such as those involving high-$T_C$ superconductors and Majorana fermions[1]. As funding and hype in fields like quantum technologies ramp up, the risk increases that scientific scandals will spread. What happens to public trust in science when breakthroughs are later debunked or technologies fail to materialize? If our research becomes plagued by replication crises and we lose the public's trust in the robustness of science, we fear that our physics futures will be in peril.

At the same time, in our experience, many physicists argue that the replication crisis is not a problem in their domains. They claim that hard-science fields are generally self-correcting, due to their cumulative knowledge building, distributed lab-bench experiments and often close proximity to commercialization, where practical applications provide a tangible test of success or failure[2].

But think again. How often would a physicist with the relevant expertise and tools who strictly follows published research methods reach the same conclusions as a given paper? What if they try to re-analyse the raw data published alongside a paper, following the analysis methods? These are questions of replicability and reproducibility, and their relevance is not limited to a few 'problem' areas[3,4].

Ensuring proper replicability requires developing safeguards. It's tempting to approach the problem as a technical one: if the community instigates the right checklists, replication issues will be solved. However, the problem is also social. Physicists face pressure from funding bodies, universities and peers to rapidly publish bold claims in high-impact journals. This pressure is interwoven with other norms and practices. Such norms may include basing studies on a small number of experimental samples which require highly customized setups to synthesize or measure; raw data, code, or processed data that are embargoed or not shared transparently; or the necessity for scientists to employ judgements in selecting and processing relevant datasets. What is important to note is that these norms are not misconduct: they are standard practice.

However, the very notions of replicability and objectivity are emergent and evolving, shaped by the collective actions and interactions of scientific communities. Finding ways to ensure replicability in a complex research environment requires an understanding of where scientific norms come from, and how they can be changed to meet future needs.

**Scientific knowledge as an emergent property**

In the 1970s, John Ziman, a condensed matter theorist turned sociologist of science, drew on the work of Philip Anderson — his contemporary and colleague — to extend the concept of emergence to scientific knowledge. Anderson, in his essay "More Is Different", argued that although all systems in nature are governed by universal laws, certain properties, such as those resulting from spontaneous symmetry breaking, are emergent[5]. These emergent properties cannot be fully predicted based on fundamental laws at smaller scales but arise instead from the collective behaviour of a system's components.

In a similar manner, argued Ziman, what counts as valid, reproducible knowledge in one science emerges from the set of norms and practices associated with that science[6]. Ziman warned against idealizing scientific norms as static and universal. Instead, he argued that understanding how these norms are negotiated in response to external pressures is key to addressing issues like the replication crisis. This view resonates with developments in the history and philosophy of science, where it is acknowledged that scientific norms, including the very notion of 'objectivity', are subject to historical change.

**Objectivity and replicability as evolving norms**

The concept of objectivity itself has evolved significantly over time, as documented by historians of science such as Lorraine Daston and Peter Galison in their book *Objectivity*. They trace how the understanding of what it means to be an 'objective' scientist has shifted through various historical eras, affecting both the training of scientists and the perception of what constitutes good science[7]. Initially, in the 18th century, the ideal was truth-to-nature, where scientists were expected to refine observations to represent the ideal forms of phenomena. By the mid-19th century, this gave way to mechanical objectivity, which sought to eliminate personal interpretation entirely, relying instead on instruments and unbiased recording methods like photography. Finally, the 20th century brought the ideal of trained judgement, blending objectivity with expert discernment to interpret complex data. These shifts in the meaning of objectivity show that what counts as rigorous science is historically contingent and culturally shaped.

These historical transformations in objectivity resonate strongly with the replication crisis in modern science. The replication crisis underscores the fact that replicability, often considered a hallmark of objective science, is itself a concept that varies according to which version of objectivity is in play. When mechanical objectivity dominated, replication meant exact duplication of a result under identical conditions, ensuring that findings were free from subjective interpretation. This notion, however, is often unrealistic in complex fields such as biology and psychology, where exact replication is difficult due to variations in experimental conditions, human subjects, and statistical methods.

The failure of many scientific studies to replicate today suggests that the scientific community's understanding of replicability may need to evolve again, just as it did for objectivity. Trained judgement might offer a more nuanced perspective, acknowledging that some level of expert interpretation and context-specific adjustment is often necessary to achieve meaningful reproducibility. However, trained judgement, if not transparently communicated, can also open the door to confirmation bias and selective reporting, which are among the very issues driving the current crisis.

Feminist philosopher of science Helen Longino provides an additional layer of analysis, arguing for a social model of objectivity that shifts focus from individual scientists to the community as a whole[8] . Longino emphasizes that objectivity is not achieved by eliminating personal bias through detachment but by promoting critical interaction and diverse perspectives within scientific communities. This approach positions the replication crisis not simply as a technical problem of flawed methodology but as a sociological issue — a failure of the scientific community to maintain sufficient diversity of viewpoints and openness to scrutiny.

This perspective aligns with historian of science Naomi Oreskes, who has argued that scientific knowledge is inherently social, shaped by collective practices and continually refined as new perspectives enter the field. For Oreskes, scientific knowledge is always provisional, built through processes of debate, critique and consensus-building[9] . Thus, a crisis of replication is also a crisis of scientific communication and community norms. When the community norms discourage dissent, prioritize novel findings over careful replication, or impose rigid publication requirements, even the best scientific methods can produce unreliable results.

This understanding reframes the replication crisis as a crisis of scientific culture rather than merely a technical problem to be solved by better statistics or experimental protocols. If objectivity itself is an emergent property of the interactions within a scientific community, then addressing the replication crisis requires more than enforcing stricter methodologies. It involves cultivating a culture where replication, critique, and transparency are valued as much as discovery and innovation.

**Where do we go from here?**

A step forward in this discussion requires accepting that that the knowledge we produce is moulded by the incentives, pressures, values and technologies of the day. The ties between commercialization, geopolitical motivations and basic academic research — which existed in Ziman's day — are strengthening at what appears to be an ever-increasing rate. National funding priorities and international trends shape what types of research is pursued and what questions are asked. Easily quantifiable metrics such as papers published, citations and impact factors, now critical for career advancement, influence the depth and scope of research pursued in papers. The rapid communication made possible by the internet has begun to shift the norms of research, affecting expectations about sample sizes, methodological rigour, and transparency of raw data.

We challenge physicists to reflect on the pressures shaping our current medium, define what replication and self-correction mean in this context, and re-imagine how our norms and scientific practices could respond to these pressures. For example, the traditional scientific paper format has remained largely unchanged despite the rise of the internet and social media, which allow for rapid communication and greater data sharing. This persistence highlights that the issue is not merely one of technological capability, but a deeper attachment to the paper as a closed and self-contained unit of knowledge. These values — rooted in tradition and the perception of science as a series of discrete and static results — stand in contrast to the evolving, collaborative, and interconnected nature of contemporary research. How could these technologies facilitate more rigorous evaluation and dissemination of data, code, and scientific results? Could papers become more 'living' documents or discussions with transparent histories, to allow conclusions to be updated and revised as new information comes to light? Additionally, how could replication work be published or recognized? As early career researchers, we ourselves are inspired by the work of many before us, and by the many conversations and creative solutions that are being explored by scientists and that can seed new changes[4].

We also encourage multidisciplinary collaborations and conversations between scientists, historians, sociologists and philosophers of science[3,10]. These exchanges will not only allow scientists to work more intentionally within the complex social landscape of research, but also provide new ideas and inspiration for social sciences and humanities colleagues.

At times, it may seem impossible to intervene in systems that, being the result of global financial, economic and social trends, feel too large and entrenched to change. But as scientists, we each still play our own small role in creating and reproducing the norms and practices of our field, just as we create and reproduce the small 'bits' of research, that together, will emerge as scientific knowledge in the future.